# Global training and the collaborative structure of elite U.S. science

Erjia Yan[1], Chaoqun Ni[2], Xiang Zheng[3]

[1]https://orcid.org/0000-0002-0365-9340
[2]https://orcid.org/0000-0002-4130-7602
[3]https://orcid.org/0000-0002-6619-5504

## Significance statement

Debates over scientific competitiveness often emphasize the supply of international talent, but less is known about how globally trained faculty are positioned within the U.S. research system. This study shows that foreign-degree faculty are disproportionately represented in elite scientific output. The evidence provides little support for the idea that this pattern reflects unusually distinctive topical specialization; instead, it points to their concentration in the institutions, fields, and collaborations through which highly cited science is produced. The findings suggest foreign degree training as part of the organizational infrastructure of U.S. science and highlight how international training pipelines, research-intensive institutions, and mixed domestic–foreign teams jointly shape elite scientific production.

## Abstract

Globally trained scientific labor is a substantial component of U.S. universities, yet the organizational mechanisms linking foreign degree training to elite scientific output remain poorly understood. We link comprehensive U.S. faculty rosters to more than 12 million OpenAlex-indexed faculty-publication observations from 2011 to 2020. Faculty with non-U.S. degrees constitute one-tenth of the U.S. professoriate but account for larger shares of total publications and top-1% cited papers. This overrepresentation is concentrated in high-output disciplinary domains and research-intensive institutions. Within institution × domain × rank × year strata, however, differences in top-1% output, FWCI, and corresponding-author share attenuate sharply, indicating that much of the aggregate pattern reflects organizational placement rather than large within-context citation advantages. Collaboration structure further differentiates foreign- and domestically trained faculty: mixed domestic–foreign faculty teams exhibit substantially elevated elite-output rates, and the association attenuates strongly after accounting for team size, suggesting that collaboration scale is central to the pattern. Topic-distinctiveness analyses show little evidence that foreign-degree faculty occupy unusually rare research niches. Overall, foreign-degree training is best understood less as an individual productivity attribute than as a structural feature of elite U.S. science, operating through institutional concentration and collaborative integration.

## Introduction

The contemporary U.S. research system relies heavily on globally trained scientific labor (1-3). Faculty whose terminal degrees were earned outside the United States hold a substantial share of academic positions and are especially visible in science, engineering, and other research-intensive fields (4). At the same time, policy debates over immigration, scientific competitiveness, and global talent flows have intensified interest in how internationally trained researchers contribute to the U.S. knowledge economy (5). Despite this interest, the mechanisms linking foreign-degree training to scientific impact remain insufficiently understood.

Prior research has documented the importance of international mobility and foreign-born scientists for innovation, collaboration, and high-technology sectors (6-10). Historical accounts emphasize the contributions of internationally trained scientists to major scientific and technological advances, and contemporary studies suggest that global mobility expands collaborative networks, broadens exposure to diverse ideas, and increases access to international research communities (11, 12). At the faculty level, some studies report that foreign-trained academics publish more or devote more time to research than domestic counterparts (13, 14). Yet these comparisons are difficult to interpret because the U.S. scientific system is highly stratified. Research output and citation impact are concentrated in a relatively small set of institutions, domains, and collaborative networks. Apparent advantages associated with foreign-degree training may therefore reflect where globally trained faculty are positioned, rather than uniform individual-level productivity differences.

Modern science is also increasingly organized around teams and networked institutions (15-18). Research-intensive universities produce disproportionate shares of highly cited work, and collaboration networks can shape visibility, resources, and access to high-impact research opportunities (1, 19). Foreign-degree faculty may contribute disproportionately to elite science because they are selectively integrated into these high-capacity institutional and collaborative environments. Under this interpretation, foreign-degree training is less a stand-alone individual attribute than a marker of how scientific labor is sorted into the organizational structure of elite research.

Existing evidence is not sufficient to distinguish these possibilities. Prior studies often rely on limited disciplinary samples, individual publication counts, or single-institution data (14, 20). Few studies can simultaneously measure workforce representation, institutional concentration, collaboration structure, citation impact, and topical specialization at national scale. Nor is it clear whether foreign-degree faculty contribute disproportionately because they work in different research niches or because they are embedded in the same elite scientific systems through different organizational pathways.

We address these questions by linking U.S. faculty rosters from the Academic Analytics Research Center (AARC) (21) to OpenAlex-indexed publications (22) from 2011 to 2020. The resulting panel contains more than 12 million faculty-publication observations annotated by degree origin, discipline, institution, rank, authorship role, publication impact, and collaboration structure. We use “foreign-degree faculty” to refer to faculty whose terminal degree was earned outside the United States. We ask four questions. First, are foreign-degree faculty disproportionately represented in overall and elite U.S. scientific output? Second, how much of this pattern reflects disciplinary and institutional concentration? Third, do mixed domestic–foreign teams and collaboration networks help explain elite output? Fourth, do foreign-degree faculty occupy more distinctive or niche topical spaces?

The results support a structural and collaborative interpretation. Foreign-degree faculty are overrepresented in total publications and top-1% cited papers, especially in STEM domains and research-intensive institutions. Nested models, sequential decompositions, and matched within-stratum comparisons show that much of the aggregate elite-output gap is associated with institutional and disciplinary placement. Mixed domestic–foreign teams exhibit elevated elite-output probabilities, and team-size adjustment indicates that collaboration scale is a central part of the mechanism. Topic-distinctiveness analyses provide little evidence that foreign-degree faculty disproportionately occupy rare topics. The findings suggest that globally trained faculty contribute to elite U.S. science through structural integration into the institutional and collaborative architecture of high-impact research.

## Results

### Foreign-degree faculty are disproportionately represented in elite U.S. scientific output

Figure 1 establishes the central empirical pattern. Faculty with non-U.S. degrees constituted approximately 11 to 12% of the U.S. professoriate during 2011 to 2020, but they consistently accounted for larger shares of total publications and top-1% cited papers. The disparity is larger at the upper tail of the citation distribution than for publications overall, yielding elite dependency ratios of approximately 1.3 to 1.5 across the decade. Thus, faculty trained outside the United States are disproportionately represented in the upper tier of U.S. scientific production.

The corresponding-author analysis suggests that the pattern is not limited to broad coauthorship participation. Foreign-degree faculty are modestly overrepresented among corresponding authors overall, but more strongly overrepresented among corresponding-author papers in the top-1% citation tier. These patterns motivate the remaining analyses, which examine whether the aggregate overrepresentation is associated with institutional sorting, collaboration structure, or topical specialization.

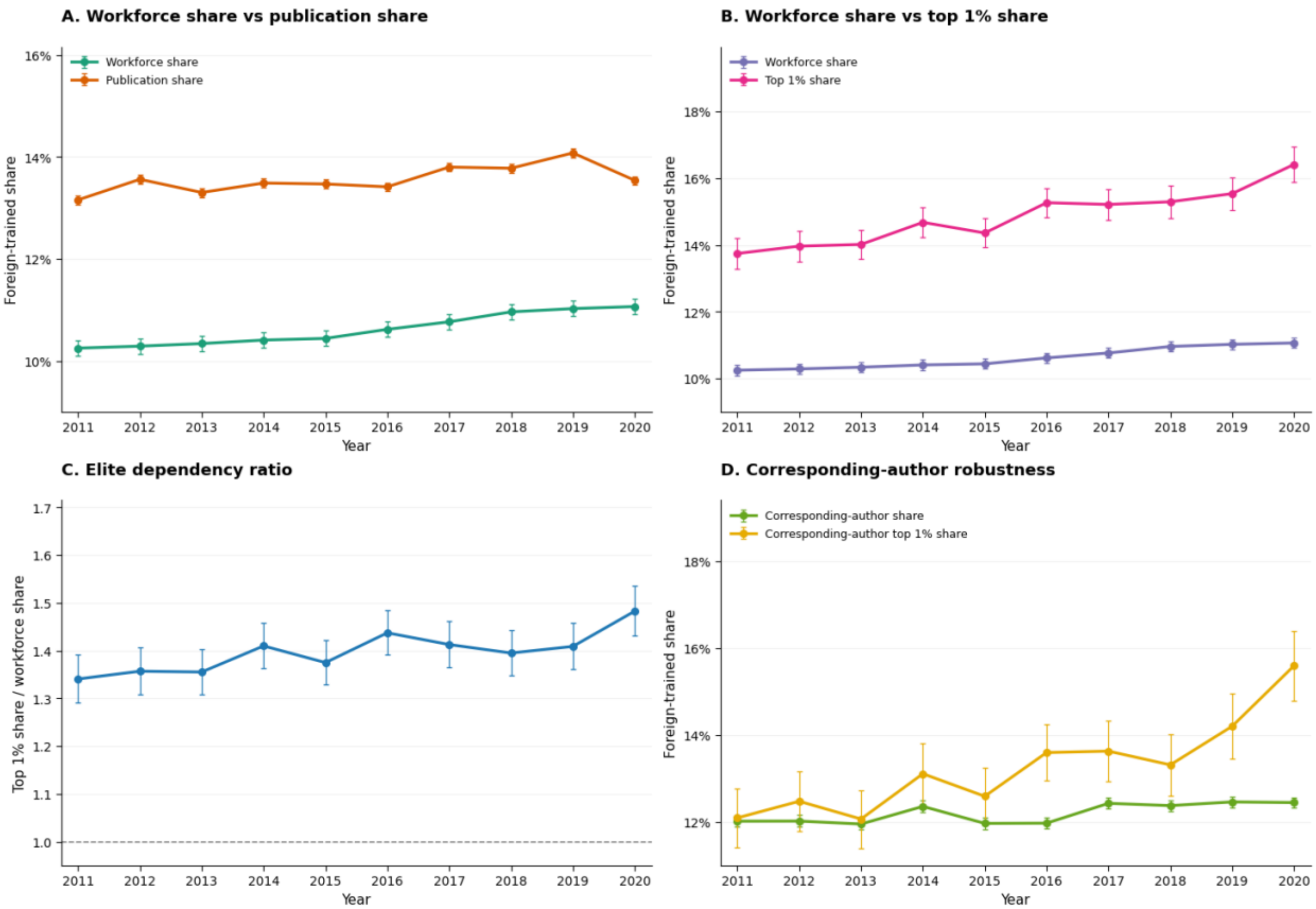

**Fig. 1 | Structural representation of foreign-degree faculty in U.S. scientific production, 2011–2020.** (A) Faculty with non-U.S. degrees constituted approximately 11-12% of the U.S. professoriate but accounted for a larger share of total publications. (B) Their representation was higher still among top-1% cited papers, exceeding workforce share in every year. (C) The elite dependency ratio, defined as the foreign-degree share of top-1% cited papers divided by the foreign-degree share of faculty, remained persistently above 1.0. (D) Restricting to corresponding-author roles produced qualitatively similar patterns. Error bars indicate approximate 95% confidence intervals derived from annual aggregate counts. Publication and top-1% shares are computed using fractional group allocation: mixed domestic–foreign faculty papers contribute one-half to each group.

## Structural concentration of foreign-degree faculty in elite scientific environments

Figure 2 shows that foreign-degree overrepresentation in elite U.S. science is concentrated within specific disciplinary and institutional segments of the research hierarchy. Foreign-degree faculty are not evenly distributed across the academic system. They are concentrated in disciplinary domains and institutional settings that produce disproportionate shares of elite scientific output, particularly STEM domains and institutions with the highest research activity. Both public and private nonprofit institutions show foreign-degree overrepresentation in elite output relative to faculty composition.

Nested models reinforce this interpretation. The raw foreign-degree association with top-1% output is positive but modest (OR = 1.03, 95% CI [1.02, 1.03]) and attenuates after incorporating disciplinary and institutional controls (fully adjusted OR approximately 1.02). The sequential Fairlie decomposition in the inset shows that the foreign–domestic top-1% publication gap narrows after disciplinary and institutional blocks are introduced, whereas rank/career-stage composition contributes little additional attenuation.

Matched within-stratum analyses provide a more direct comparison among faculty working in similar institutional and disciplinary contexts. Within institution × domain × rank × year strata containing both groups, the raw top-1% publication gap (+0.09 papers per faculty-year) collapses to near parity (+0.004), and differences in field-weighted citation impact (FWCI) and corresponding-author share similarly approach zero. Publication volume (+0.89 publications per faculty-year) and mixed-team participation (+0.27 share points), however, remain higher for foreign-degree faculty after matching. This pattern indicates that much of the aggregate elite-output difference is associated with organizational placement, while collaboration remains a persistent distinguishing feature among otherwise comparable faculty.

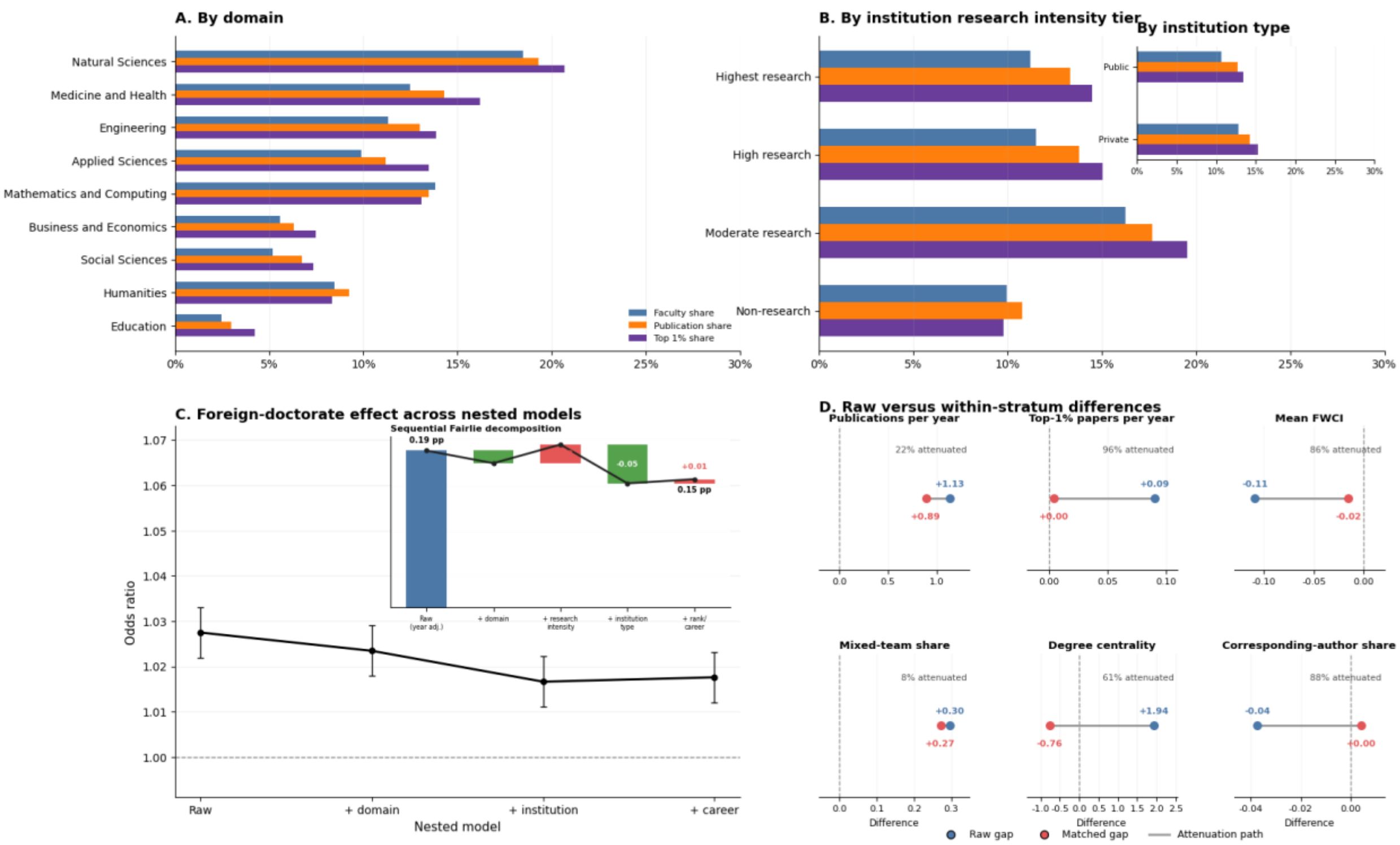

**Fig. 2 | Structural concentration of foreign-degree faculty within the U.S. research system.** (A) Foreign-degree faculty are unevenly distributed across disciplinary domains, with higher representation in research-intensive STEM areas. (B) Foreign-degree representation varies by institutional research intensity, with the strongest concentration in highest- and high-research institutions. The inset reports institutional type. (C) Nested logistic regressions show attenuation of the foreign-degree association after adjustment for disciplinary composition, institutional context, and career stage. Odds ratios are reported with 95% confidence intervals. The inset presents a sequential Fairlie decomposition of the foreign–domestic top-1% publication gap. (D) Raw versus matched within-stratum differences across faculty-level

outcomes. Matched comparisons are computed within institution × domain × rank × year strata containing both groups. Points denote raw and matched foreign–domestic differences; connecting lines show attenuation.

**Mixed domestic–foreign teams and network embeddedness are associated with elite scientific output**

The most pronounced collaboration pattern is the elevated elite-output rate associated with mixed domestic–foreign teams (Panel A). Among papers with AARC-linked faculty authors, those involving both domestic- and foreign-trained faculty are substantially more likely to enter the top-1% citation tier than papers produced by homogeneous teams. This pattern persists after adjusting for disciplinary composition and institutional setting, indicating that mixed-team overrepresentation is not reducible solely to concentration within elite scientific environments (Panel B).

At the same time, the attenuation pattern in the nested models suggests that a substantial portion of the mixed-team association is accounted for by collaboration scale. Incorporating team size reduces the mixed-team odds ratio from approximately 1.76 to 1.10, implying that the observed association partly reflects the tendency of mixed teams to participate in larger collaborative structures. Nevertheless, even after accounting for collaboration scale, mixed teams retain a positive association with elite output. The sharp attenuation after team-size adjustment indicates that collaboration scale is central to the mixed-team pattern; the remaining positive association suggests that mixed teams may also capture other features of collaborative integration.

Descriptive model-implied curves suggest that the foreign–domestic elite-output gap widens with collaboration-network centrality, especially among highly connected faculty (Panel C). As network centrality increases, foreign-trained faculty exhibit progressively larger predicted elite-output returns relative to domestically trained faculty. The divergence is modest at low levels of embeddedness but widens systematically among highly connected researchers, indicating that network position is converted into elite scientific impact more strongly for foreign-trained faculty. Combined with the structural and matched analyses, these findings are consistent with the interpretation that the contribution of foreign-degree faculty to elite U.S. science emerges through collaborative integration within large-scale research networks rather than isolated individual-level differences alone.

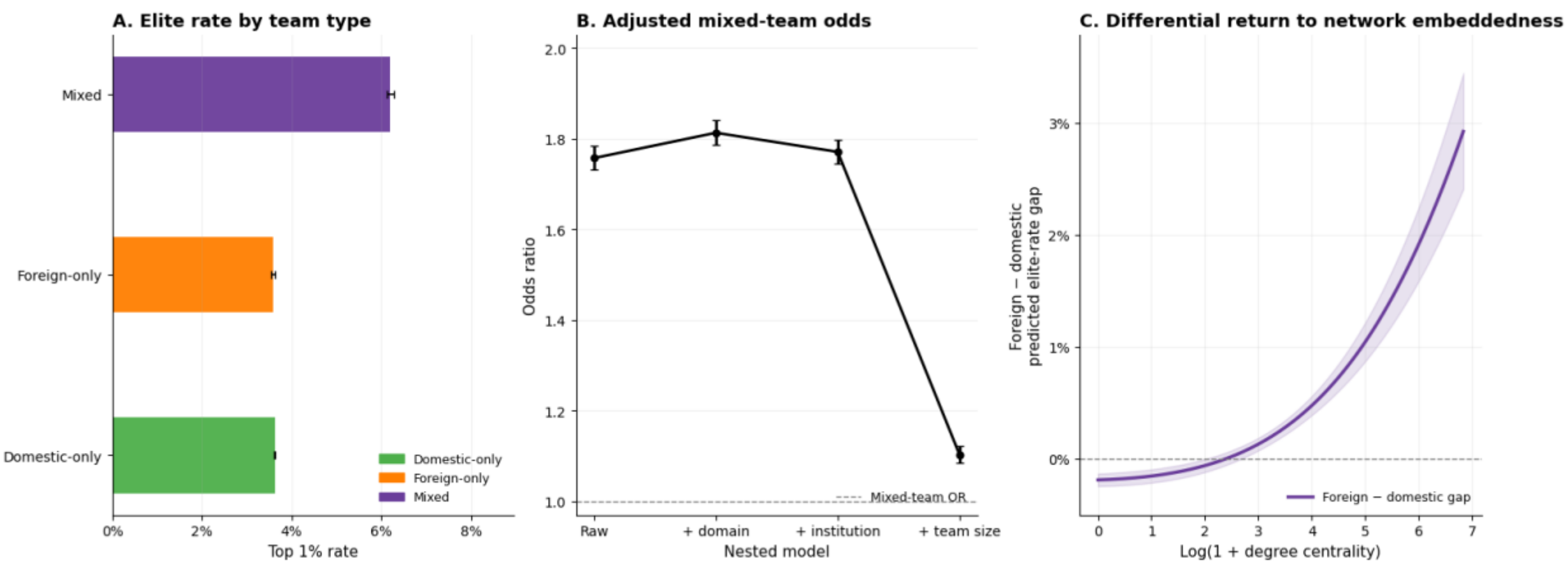

**Fig. 3 | Collaborative mechanisms underlying elite scientific output among papers with AARC-linked faculty authors.** (A) Elite-output rates differ by team composition, with mixed domestic–foreign faculty teams exhibiting the highest top-1% rate. Error bars denote 95% confidence intervals based on Wilson intervals. (B) Nested logistic regressions show that the mixed-team association remains elevated after domain and institutional adjustment but attenuates after accounting for team size. Odds ratios are relative to domestic-only teams. (C) Differential return to network embeddedness from a descriptive model of elite-output probability across degree centrality. Positive values indicate larger predicted foreign–domestic gaps at higher centrality; shaded regions denote approximate 95% confidence intervals.

## Foreign-degree faculty do not occupy more distinctive research niches

Figure 4 assesses a plausible alternative explanation: foreign-degree faculty may appear overrepresented in elite output because they work in especially rare or distinctive research niches. The results provide little support for that account. Across assistant, associate, and full professors, foreign- and domestically trained faculty exhibit highly overlapping topic-distinctiveness distributions. Year-adjusted foreign–domestic gaps are very small in magnitude: about 0.09% for assistants, -0.43% for associates, and -0.09% for full professors relative to the rank-specific mean, and are slightly negative for two of the three ranks.

The distributional evidence points in the same direction. Foreign-degree faculty do not disproportionately occupy the high-distinctiveness tail, where unusually rare topic portfolios would be concentrated. Comparisons across the largest foreign-degree origin countries likewise show tightly clustered distinctiveness estimates near the domestic benchmark, with no major origin system standing out as systematically more distinctive. These results make it unlikely that the elite-output advantage of foreign-degree faculty is explained primarily by topical rarity or niche specialization. The more plausible mechanisms are organizational and collaborative: concentration in high-capacity research environments and sustained participation in mixed domestic–foreign collaboration networks.

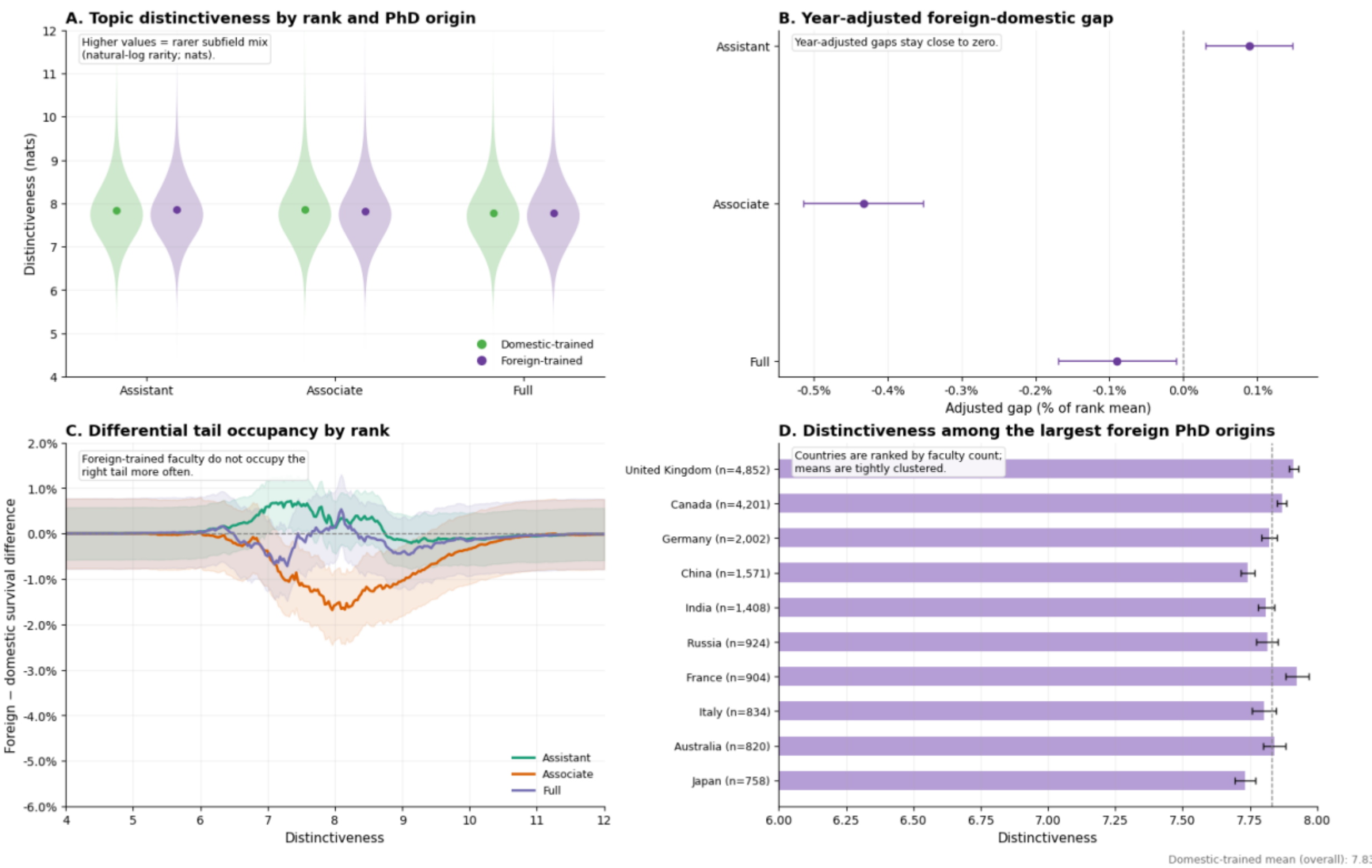

**Fig. 4 | Topic distinctiveness does not explain the elite-output overrepresentation of foreign-degree faculty.** (A) Topic-distinctiveness distributions are highly similar for foreign- and domestically trained faculty across academic ranks. (B) Year-adjusted foreign–domestic gaps are close to zero across ranks. (C) Tail-occupancy curves show that foreign-degree faculty do not disproportionately occupy the high-distinctiveness tail. (D) Distinctiveness values are tightly clustered across the largest foreign-degree origin countries. Topic distinctiveness is measured as an information-theoretic rarity index derived from annual OpenAlex topic distributions; higher values indicate rarer topic portfolios relative to the annual U.S. faculty baseline.

## Heterogeneity and robustness checks

Supplementary career-stage and cohort analyses show that foreign-degree overrepresentation in productivity, elite output, and mixed-team participation is not confined to a single career stage or cohort (Supplementary Fig. 1). Foreign-degree faculty publish more papers per faculty-year, participate more often in mixed domestic–foreign collaborations, and produce more top-1% papers per faculty-year across career stages, years since degree, and cohorts. The magnitude of the gap varies across career and cohort bins, but the pattern is broadly consistent with possible cumulative advantage, selective retention, and long-term integration into collaborative scientific networks rather than a narrow entry-stage effect alone. These cohort patterns do not condition jointly on institution, domain, rank, and year; the matched analyses in Figure 2D show that elite-output differences attenuate sharply when those contextual factors are held fixed.

Country-of-training analyses indicate that the foreign-degree contribution to elite U.S. science is broadly distributed across multiple major degree-origin systems rather than driven by a single dominant country (Supplementary Fig. 2). United Kingdom- and Canada-trained faculty constitute the largest groups among foreign-degree origin faculty, whereas China-trained faculty account for disproportionately large shares of publications and top-1% papers relative to their representation among foreign-degree origin authors. Across nearly all major origin systems, foreign-degree faculty are heavily concentrated in highest-research U.S. institutions and exhibit high levels of mixed domestic–foreign collaboration. Topic distinctiveness remains tightly clustered across countries, suggesting that country-level heterogeneity in elite-output representation is not primarily a function of unusual topical specialization.

Supplementary robustness checks support the main conclusions. The proportional benchmark indicates that observed foreign-degree participation exceeded workforce-share expectations by approximately 900 top-1% papers per year, corresponding to more than 9,000 excess elite publications over 2011 to 2020 (Supplementary S7). Parallel upper-tail models show that the foreign-degree association persists across top-10% and top-1% thresholds, and continuous-impact models based on log(FWCI) indicate a modest upward shift after structural controls (Supplementary S8). Institution-level panel models show that elite output rises with foreign-degree faculty share but at diminishing rather than accelerating returns (Supplementary S9).

## Discussion

This study provides a large-scale account of how foreign-degree faculty participate in elite U.S. science. Although they constitute roughly one-tenth of the U.S. professoriate, foreign-degree faculty account for larger shares of publications and top-1% cited papers. This overrepresentation is persistent across 2011 to 2020 and is especially visible in the upper tail of the citation distribution. The result indicates that globally trained faculty are not peripheral to U.S. academic science; they are disproportionately present in the system of elite scientific production.

The evidence also cautions against interpreting this pattern as a simple individual-level productivity premium. Much of the aggregate overrepresentation is associated with where foreign-degree faculty work: research-intensive institutions, high-output domains, and collaborative environments that generate disproportionate shares of elite science. Matched within-stratum comparisons show that top-1% output, FWCI, and corresponding-author differences attenuate substantially when faculty are compared within the same institution, domain, rank, and year. The observed pattern is therefore better understood as a feature of the organization of scientific labor than as evidence of a large intrinsic difference between foreign- and domestically trained faculty operating under equivalent conditions.

A second central pattern concerns collaboration. Mixed domestic–foreign teams exhibit elevated elite-output rates, although this association attenuates sharply after adjustment for team size, suggesting that such teams are embedded within the large collaborative structures most strongly associated with elite science. Foreign-degree faculty also participate in mixed domestic–foreign collaborations at substantially higher rates across career stages, cohorts, and major countries of degree origin. The network analyses reinforce this interpretation: foreign-trained faculty exhibit larger elite-output returns to collaboration-network embeddedness, particularly at higher levels of degree centrality, and descriptive patterns indicate widening

foreign–domestic gaps among highly connected researchers. These findings suggest that the contribution of globally trained faculty emerges less through isolated individual productivity differences than through integration into large-scale collaborative and institutional infrastructures that disproportionately generate elite science.

Equally important, the topic-distinctiveness analyses provide little evidence that foreign-trained faculty contribute disproportionately through systematically more novel or specialized research agendas. Across assistant, associate, and full professor ranks, foreign- and domestically trained faculty exhibit highly similar distinctiveness distributions. Year-adjusted differences are uniformly small, foreign-trained faculty are not overrepresented in the extreme right tail of the distribution, and average distinctiveness values remain tightly clustered across major countries of degree origin. The advantage associated with foreign-degree faculty therefore appears to arise less from exceptional topical differentiation than from organizational and collaborative positioning within the research system.

These findings have broader implications for understanding the organization of contemporary science (11). The results suggest that international scientific mobility contributes to national research capacity not merely through the addition of individual talent, but through the integration of globally trained researchers into collaborative and institutional systems associated with elite scientific production (12, 23). In this sense, foreign-trained faculty represent an infrastructural component of the U.S. scientific enterprise (24). Policies affecting international mobility, faculty recruitment, and long-term retention may therefore influence not only workforce composition but also the collaborative architecture through which elite scientific output is produced (5, 25).

Several limitations should be noted. First, our analysis is observational and cannot fully separate the effects of foreign-degree training from selection into the U.S. academic system. Foreign-trained faculty who obtain U.S. faculty positions may represent positively selected subsets of international researcher populations (26). Second, the AARC faculty census primarily captures research-oriented academic institutions, and the observed patterns may differ in less research-intensive sectors. Third, our analysis focuses on degree-training origin rather than nationality or citizenship; foreign-trained faculty may include U.S.-born scholars trained abroad, while domestically trained faculty may include foreign-born scholars trained in the United States. Finally, collaboration-network measures are derived from observed coauthorship structures and cannot directly identify the underlying social or organizational processes that generate collaborative advantages. Despite these limitations, the results provide a large-scale empirical account of how globally trained scientific labor contributes to elite U.S. research. Foreign-degree faculty are not simply overrepresented as individuals; rather, they are disproportionately embedded within the institutional and collaborative structures through which elite science is produced.

## Data and methods

### Data sources and panel construction

We linked annual U.S. faculty rosters from the Academic Analytics Research Center (AARC) (21) to OpenAlex-indexed publications (22) from 2011 to 2020. AARC records provide faculty identifiers, employing institution, terminal degree institution, discipline, rank, and annual observation year. OpenAlex provides work identifiers, publication year, authorship metadata, field-weighted citation impact, and field- and year-normalized top-10% and top-1% citation indicators. Faculty were included if they appeared in an AARC annual snapshot and were linked to an OpenAlex author identifier with high or medium confidence.

Faculty-publication observations were constructed contemporaneously: publications were attributed to a faculty member only when the publication year matched the AARC observation year. This design aligns workforce composition and scientific output within the same annual observation window. The resulting panel contains more than 12 million faculty-publication observations.

### Degree origin

Degree origin was defined using the country of the terminal degree-granting institution. Faculty whose degree institution was in the United States were classified as U.S.-degree faculty; faculty whose degree institution was outside the United States were classified as foreign-degree faculty. Faculty with unresolved degree-country information were excluded from origin-specific analyses. The classification measures degree-training origin, not nationality, citizenship, or birthplace.

### Counting rules and impact measures

System-level shares use fractional group allocation to avoid double counting across degree-origin groups. Domestic-only papers were assigned fully to the domestic group; foreign-only papers were assigned fully to the foreign group; and mixed domestic–foreign papers were assigned 0.5 to each group. Faculty-year analyses use full faculty participation counts: each faculty member receives one publication count for each linked work in that faculty-year. The two counting approaches answer different questions: fractional group allocation estimates system-level origin shares, whereas faculty-year counts measure individual participation and productivity.

Elite scientific output was defined as a publication in the top 1% of the OpenAlex field- and year-normalized citation distribution. We also analyzed top 10% indicators and continuous field-weighted citation impact (FWCI) in supplementary robustness checks.

### Structural concentration, decomposition, and matching

Nested logistic regressions (27) estimated the association between degree origin and top-1% publication probability under progressively richer structural controls. Models were estimated on grouped paper-faculty observations with binomial likelihoods and robust standard errors. Specifications added year and domain, institutional research intensity and institutional control, and career stage. These models are descriptive and are used to assess attenuation rather than to estimate a causal effect of degree origin.

We implemented a sequential Fairlie-style decomposition (28) for the binary top-1% outcome. The decomposition sequentially incorporated disciplinary domain, institutional research intensity, institution type, and career stage while holding publication year fixed. For each specification, we computed standardized foreign–domestic gaps by predicting outcomes under counterfactual foreign and domestic assignments while preserving the observed covariate distribution. Because nonlinear sequential decompositions are order-dependent, the estimates are interpreted as descriptive attenuation rather than causal attribution.

Matched faculty-year comparisons were computed within institution × domain × rank × year strata. For each stratum containing both foreign- and domestically trained faculty, we computed the foreign–domestic difference in publications per year, top-1% papers per year, mixed-team share, degree centrality, corresponding-author share, and mean FWCI. The matched estimate is the average within-stratum difference across valid strata (n = 38,223).

### Mixed-team and network analyses

Each paper containing at least one AARC-linked faculty author was classified as domestic-only, foreign-only, or mixed based on the degree origin of participating faculty authors. We estimated paper-level logistic regressions of top-1% status on team type, with sequential adjustment for publication year, disciplinary domain, institutional context, and team size. Team type refers to the degree-origin composition of AARC-linked U.S. faculty authors on the paper, not the full global author list. Odds ratios are reported relative to domestic-only teams.

Faculty collaboration networks were constructed from coauthorship ties among AARC-linked faculty. Degree centrality was defined as the number of distinct faculty collaborators. Network-return curves were generated from model-implied predictions across degree centrality.

### Topic distinctiveness and falsification analyses

To evaluate whether foreign-degree faculty disproportionately occupy unusually rare or specialized research niches, we constructed an information-theoretic topic-distinctiveness measure using OpenAlex topic classifications (n = 4,516). For each year from 2011–2020, we estimated the baseline probability distribution of research topics across the U.S. professoriate. We then calculated a faculty-year distinctiveness index that measures the rarity of an author's topic portfolio relative to the annual national baseline:

$$D_i = \sum_{s} p_{is} \left[-\log(p_s)\right]$$

where:

- $p_{is}$denotes the faculty member's publication share in topic $s$,
- $p_s$denotes the annual prevalence of topics $s$within the U.S. faculty population.

The resulting index is expressed in natural-log information units ("nats"), with higher values indicating publication portfolios concentrated in rarer or more distinctive intellectual areas. By benchmarking against annual national topic distributions, the measure accounts for temporal shifts in disciplinary composition while preserving relative topical rarity within each year.

## Conflict of interest statement

The authors declare no competing interests.

## Author contributions

EY, CN, XZ: Conceptualization, Methodology, Software. EY: Data curation, EY, CN, XZ: Writing - Original draft preparation. EY: Visualization.

## AI use statement

Portions of code development and text editing were assisted using a large language model (ChatGPT 5.5). All outputs were reviewed, validated, and edited by the authors.

## References

1. J. Gu, X. Pan, S. Zhang, J. Chen, International mobility matters: Research collaboration and scientific productivity. *Journal of Informetrics* **18**, 101522 (2024).
2. R. B. Freeman, Immigration, international collaboration, and innovation: Science and technology policy in the global economy. *Innovation policy and the economy* **15**, 153-175 (2015).
3. H. De Wit, *Internationalization of higher education in the United States of America and Europe* (Information Age Pub Incorporated, 2009).
4. R. B. Freeman, Globalization of scientific and engineering talent: international mobility of students, workers, and ideas and the world economy. *Economics of Innovation and New Technology* **19**, 393-406 (2010).
5. M. A. Barteau, S. M. Rovito, *International talent programs in the changing global environment* (National Academies Press, 2024).
6. T. Yuret, An analysis of the foreign-educated elite academics in the United States. *Journal of Informetrics* **11**, 358-370 (2017).
7. K. Mamiseishvili, Characteristics, job satisfaction, and workplace perceptions of foreign-born faculty at public 2-year institutions. *Community College Review* **39**, 26-45 (2011).
8. T. A. Velema, The contingent nature of brain gain and brain circulation: Their foreign context and the impact of return scientists on the scientific community in their country of origin. *Scientometrics* **93**, 893-913 (2012).
9. P. E. Stephan, S. G. Levin, Exceptional contributions to US science by the foreign-born and foreign-educated. *Population research and Policy review* **20**, 59-79 (2001).
10. D. Kim, S. B. Twombly, L. Wolf-Wendel, A. A. Belin, Understanding career mobility of professors: Does Foreign-born status matter? *Innovative Higher Education* **45**, 471-488 (2020).
11. C. S. Wagner, L. Leydesdorff, Network structure, self-organization, and the growth of international collaboration in science. *Research policy* **34**, 1608-1618 (2005).
12. C. Franzoni, G. Scellato, P. Stephan, Foreign-born scientists: mobility patterns for 16 countries. *Nature biotechnology* **30**, 1250-1253 (2012).
13. J. J. Lee, C. Rice, Welcome to America? International student perceptions of discrimination. *Higher education* **53**, 381-409 (2007).
14. K. Mamiseishvili, V. J. Rosser, International and citizen faculty in the United States: An examination of their productivity at research universities. *Research in Higher Education* **51**, 88-107 (2010).
15. S. Wuchty, B. F. Jones, B. Uzzi, The increasing dominance of teams in production of knowledge. *Science* **316**, 1036-1039 (2007).
16. S. M. Fiore, Interdisciplinarity as teamwork how the science of teams can inform team science. *Small Group Research* **39**, 251-277 (2008).

17. L. Wu, D. Wang, J. A. Evans, Large teams develop and small teams disrupt science and technology. *Nature* **566**, 378-382 (2019).
18. B. F. Jones, S. Wuchty, B. Uzzi, Multi-university research teams: Shifting impact, geography, and stratification in science. *science* **322**, 1259-1262 (2008).
19. V. Larivière, Y. Gingras, C. R. Sugimoto, A. Tsou, Team size matters: Collaboration and scientific impact since 1900. *Journal of the Association for Information Science and Technology* **66**, 1323-1332 (2015).
20. K. L. Webber, Research productivity of foreign-and US-born faculty: Differences by time on task. *Higher Education* **64**, 709-729 (2012).
21. G. Tripodi *et al.*, Tenure and research trajectories. *Proceedings of the National Academy of Sciences* **122**, e2500322122 (2025).
22. J. Priem, H. Piwowar, R. Orr, OpenAlex: A fully-open index of scholarly works, authors, venues, institutions, and concepts. *arXiv preprint arXiv:2205.01833* (2022).
23. C. R. Sugimoto *et al.*, Scientists have most impact when they're free to move. *Nature* **550**, 29-31 (2017).
24. P. Stephan, *How economics shapes science* (Harvard University Press, 2015).
25. W. R. Kerr, W. F. Lincoln, The supply side of innovation: H-1B visa reforms and US ethnic invention. *Journal of Labor Economics* **28**, 473-508 (2010).
26. G. J. Borjas (1987) Self-selection and the earnings of immigrants. (National Bureau of Economic Research).
27. D. W. Hosmer Jr, S. Lemeshow, R. X. Sturdivant, *Applied logistic regression* (John Wiley & Sons, 2013).
28. R. W. Fairlie, An extension of the Blinder-Oaxaca decomposition technique to logit and probit models. *Journal of economic and social measurement* **30**, 305-316 (2005).
29. X. Zheng, C. Ni, The tenure debate: how US faculty change their research practices post-tenure. *iConference 2024 Proceedings* (2024).
30. American Council on Education (2026) 2025 Institutional Classification.
31. J. J. Heckman, E. Leamer, Handbook of econometrics. (2009).

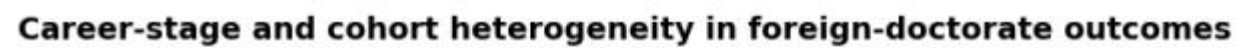

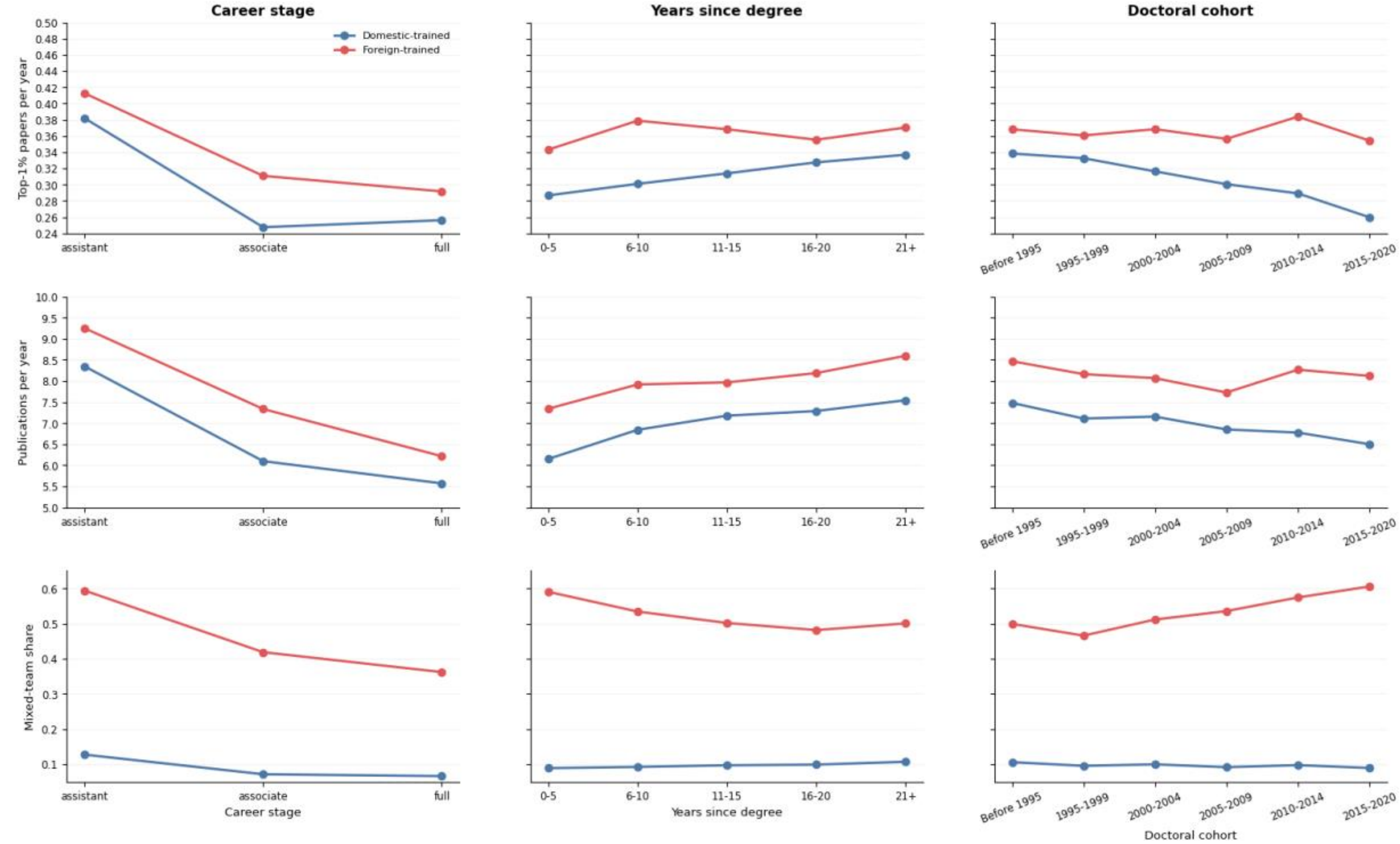

**Supplementary Fig. S1. Career-stage and cohort heterogeneity in foreign-degree outcomes.** Foreign- and domestically trained faculty are compared across career stage, years since degree, and cohort for three core outcomes: top-1% papers per faculty-year, publications per faculty-year, and mixed-team participation. Across nearly all career stages, experience bins, and cohorts, foreign-degree faculty exhibit higher annual publication volume, higher top-1% paper production, and substantially greater participation in mixed domestic–foreign collaborations. The foreign–domestic advantage in top-1% papers per faculty-year persists throughout the career distribution, although its magnitude varies across cohorts and experience levels. These patterns suggest that the disproportionate contribution of foreign-degree faculty to elite U.S. science reflects sustained productivity and collaborative integration across academic careers rather than a narrow advantage confined to a single career stage or cohort.

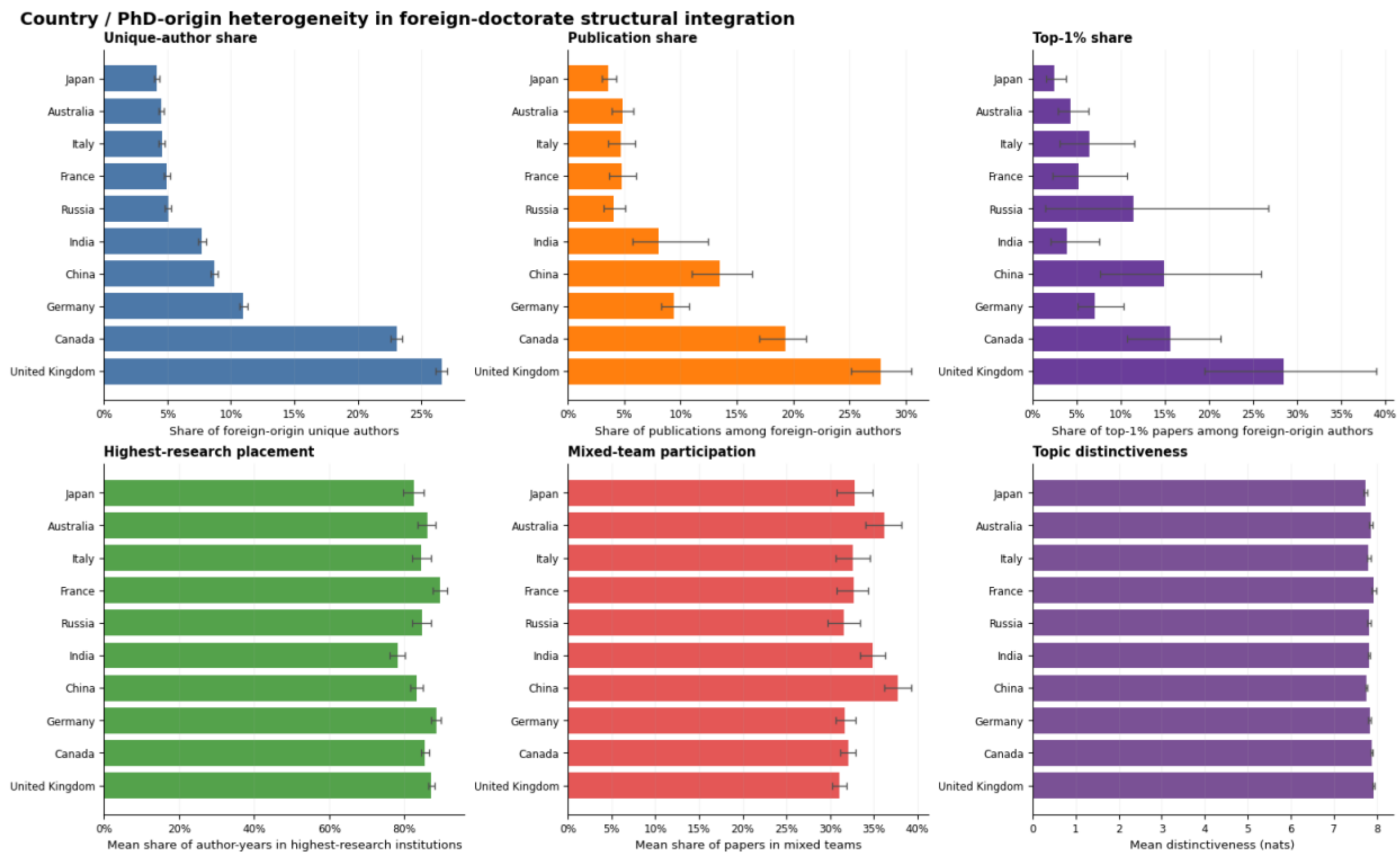

**Supplementary Fig. S2. Country / Degree-origin heterogeneity in foreign-degree structural integration.** The largest foreign-degree origin countries are compared using unique authors as the analytic unit. Panels report each country's share of foreign-degree origin authors, publication share, top-1% paper share, concentration in highest-research U.S. institutions, mixed domestic–foreign collaboration rates, and topic distinctiveness. Error bars denote 95% bootstrap confidence intervals. Considerable heterogeneity exists across origin systems: United Kingdom- and Canada-trained faculty constitute the largest foreign-degree origin groups, while China-trained faculty account for disproportionately large shares of publications and top-1% papers relative to their representation among foreign-degree origin authors. Nevertheless, nearly all major origin systems exhibit strong concentration in research-intensive institutional environments and high levels of mixed-team participation. Topic-distinctiveness estimates remain comparatively tightly clustered across countries, indicating that differences in elite-output representation are not primarily driven by specialization in unusually distinctive research topics.

# Supplementary materials and methods

## S1. Data sources

This study integrates administrative faculty records with a large-scale bibliometric database to construct a longitudinal panel of U.S. academic faculty and their research output from 2011–2020.

### S1.1 Academic Analytics Research Center (AARC)

Faculty employment data were obtained from the Academic Analytics Research Center (AARC), covering U.S. higher education institutions across multiple annual snapshots ("AAD versions") from 2011–2020 (21, 29). Each AAD version corresponds to a faculty census year and includes:

- Unique faculty identifiers (PersonId)
- Institutional affiliation (InstitutionId)
- Degree-granting institution (DegreeInstitutionId)
- Discipline classification (TaxonomyLevel01Id)
- Employment year (via AAD version mapping)

We developed a scalable, reproducible one-to-one entity-resolution pipeline to link 310,303 U.S.–based AARC faculty to OpenAlex (22) author IDs while minimizing false merges and missed matches within a corpus exceeding 100 million authors. We implemented surname-based blocking (exact normalized last name), reducing the search space to 2.46 million candidate pairs (~8.2 per faculty) while preserving coverage. Each candidate pair received a transparent composite score combining name agreement (surname, first-initial match, middle-initial consistency), institutional similarity (average of token-level recall and Jaccard to balance asymmetry and generic overlap), and binary discipline overlap between AARC taxonomy and OpenAlex primary field.

Candidates were deterministically ranked per faculty, and ambiguity was quantified using a separation margin ($\Delta = S_1 - S_2$), where larger margins indicate stable dominance. Empirically calibrated thresholds defined confidence tiers: High ($S \geq 0.80$ and $\Delta \geq 0.05$), Medium ($0.50 \leq S < 0.80$ and $\Delta \geq 0.01$), and Low otherwise; sensitivity tests (±0.02) altered the High tier by <3%, indicating stability. The design explicitly mitigates common-surname inflation via dominance filtering, accommodates institutional mobility and interdisciplinary breadth by weighting identity signals over field similarity, and prioritizes precision over recall; approximately 2.8% of faculty had no viable OpenAlex match, likely reflecting limited publication records or coverage gaps.

Faculty were included if they:

- Appeared in the AAD snapshot for year *t*,
- Had a degree institution identifier,
- Were successfully linked to OpenAlex with confidence tier "High" or "Medium".

Across 2011–2020, the annual faculty population ranged from approximately 146,000 to 168,000 individuals.

### S1.2 OpenAlex

Publication and citation data were obtained from the OpenAlex database (snapshot: subugoe-collaborative.openalex_walden.works) (22).

For each work, the following fields were used:

- id (OpenAlex work ID)

- publication_year
- FWCI (field-weighted citation impact)
- citation_normalized_percentile.is_in_top_10_percent
- citation_normalized_percentile.is_in_top_1_percent
- authorships (including author ID, position, corresponding flag)

Only works published between 2011 and 2020 were considered.

### S1.3 Institutional and disciplinary metadata

Institutional attributes were obtained from CCIHE (30):

- CCIHE_research2025 (research intensity level)
  - 1 = highest research intensity
  - 2 = high
  - 3 = moderate
  - -2 = non-research
- CCIHE_control
  - 1 = public
  - 2 = private nonprofit
  - 3 = for-profit

Disciplinary classifications were obtained from AARC taxonomy data. This taxonomy consists of 170 distinct disciplines. Based on disciplinary semantics, we manually aggregated these disciplines into 28 broader domains and further into 10 top-level domains. The full taxonomy and aggregation scheme are provided in Supplementary Data S1.

- Umbrella (umbrella domain)
- Area
- TaxonomyLevel01Name

Degree country was obtained and disambiguated from degree institution provided by AARC. We manually identified the country location of all degree-granting institutions, resolving ambiguities and harmonizing historical country names with present-day entities. In total, degree-granting institutions span 131 countries. Across the dataset, we identified 3,922 unique institutions, including 1,321 U.S. institutions and all 395 AARC-indexed employing institutions.

## S2. Record linkage and panel construction

For each faculty member observed in AAD year $t$, publications were attributed only if: publication_year = $t$. This ensures that research output is aligned with employment-year status, avoiding temporal leakage from prior or subsequent employment periods.

A one-time author–work spine table was constructed linking:

- Faculty-year metadata
- OpenAlex works
- Authorship roles
- Citation-normalized metrics

The spine table includes:

- Faculty metadata (Degree origin, discipline, institutional attributes)
- Work-level attributes (FWCI, top 10%, top 1%)
- Authorship position indicators

## S3. Classification of Degree origin

Faculty were classified based on terminal degree institution country:

- **Domestic-trained**: Terminal degree institution located in the United States.
- **Foreign-trained**: Terminal degree institution located outside the United States.

Degree origin was treated as time-invariant across the panel.

## S4. Publication counting and fractional allocation

For each publication in year *t*, authorship was examined to determine whether:

- domestic-trained faculty were present
- foreign-trained faculty were present

To avoid double counting mixed-authorship papers, fractional weights were applied:

- Domestic-only paper → domestic = 1
- Foreign-only paper → foreign = 1
- Mixed paper → domestic = 0.5, foreign = 0.5

This ensures: $\text{DomesticShare}_t + \text{ForeignShare}_t = 1$

## S5. Authorship roles

Authorship roles were defined using OpenAlex authorship metadata:

- **Any author**: is_author = TRUE
- **Corresponding author**: is_corresponding = TRUE

## S6. Citation-normalized impact measures

OpenAlex provides FWCI, defined as:

$$\text{FWCI} = \frac{\text{CitationsReceived}}{\text{ExpectedCitations}}$$

normalized by publication year, work type, and topic. Mean FWCI was computed using fractional weights at the paper level.

OpenAlex provides binary indicators:

- is_in_top_10_percent
- is_in_top_1_percent

These are field- and year-normalized. Shares of top-10% and top-1% papers were computed using fractional allocation relative to total elite papers in each year.

## S7. Counterfactual models

To assess the macro-level implications of elite overrepresentation, we implemented a proportional counterfactual framework (31). For each year *t*, we computed:

$$E_{obs,t} = T_t \times S_{f,t}^{elite}$$

where$T_t$ is the total number of top-1% cited papers in year *t*, and $S_{f,t}^{elite}$ is the observed foreign-trained share of elite output.

We then computed a proportional benchmark:

$$E_{cf,t} = T_t \times S_{f,t}^{faculty}$$

where $S_{f,t}^{faculty}$ is the foreign-trained faculty share in that year.

The annual elite surplus is:

$$\Delta E_t = E_{obs,t} - E_{cf,t}$$

All quantities were derived from the full publication dataset spanning 2011–2020. The simulation operates on more than 12 million faculty–publication observations, with annual elite paper counts ranging from approximately 18,000 to 26,000 per year. Across the decade, the average annual elite surplus was approximately 900 top-1% papers. The cumulative surplus exceeded 9,000 elite publications over the 2011–2020 period. Because the benchmark assumes identical productivity conditional on workforce share, the surplus captures the structural amplification of foreign-trained scholars within high-impact segments of the research system.

## S8. Distributional and upper-tail analysis

Each observation corresponds to a faculty member associated with a publication and includes field-normalized citation indicators, institutional identifiers, research intensity classification, institutional control, authorship position, and degree training origin.

Elite status was defined using OpenAlex percentile indicators:

- is_top10: top 10% of field- and year-normalized citation impact.
- is_top1: top 1% of field- and year-normalized citation impact.

In addition, the continuous field-weighted citation impact (FWCI) metric was used to assess distribution-wide impact differences. All models were estimated at the paper–faculty level.

To assess whether the foreign-trained advantage intensifies toward the extreme upper tail of the citation distribution, we estimated parallel logistic regression models for two elite thresholds:

1. $Top10_{ijt}$
2. $Top1_{ijt}$

Both models included identical controls:

- Disciplinary fixed effects (umbrella domain),
- Publication year fixed effects,
- Institution fixed effects,
- Research intensity classification,
- Institutional control,
- Corresponding-author status,
- Interaction between foreign-trained status and leadership role.

Models were estimated using BigQuery ML logistic regression with L2 regularization (λ = 0.1). Coefficients are reported as odds ratios.

We estimated a linear regression on log-transformed field-weighted citation impact:

$$\log\left(FWCI_{ijt} + 0.01\right) = \alpha + \beta Foreign_i + X_{ijt}'\gamma + \varepsilon_{ijt}$$

The same fixed effects and control structure were included as in the logistic models. The log transformation addresses skewness in citation distributions.

The quantile-based logistic models reveal modest but systematic upper-tail steepening. In the fully specified model, foreign-trained faculty exhibit 1.6% higher odds of producing top-10% cited papers (odds ratio ≈ 1.02), while the advantage increases to approximately 3.0% higher odds at the top-1% threshold (odds ratio ≈ 1.03). The near doubling of the coefficient between top-10% and top-1% indicates that the conditional advantage becomes more pronounced at the extreme upper tail.

The continuous impact model further clarifies this pattern. The coefficient on foreign-trained status in the log(FWCI) regression is approximately 0.075, corresponding to an estimated 7–8% higher normalized citation impact after controlling for discipline, institution, research intensity, governance structure, year, and leadership role. This result indicates that foreign-trained faculty exhibit a modest upward shift across the entire citation distribution, not merely at elite thresholds.

## S9. Institutional panel model

Institution-level amplification was evaluated using a panel regression of elite output per faculty on foreign-trained faculty share and its square:

$$ElitePerFaculty_{it} = \beta_1 ForeignShare_{it} + \beta_2 ForeignShare_{it}^2 + InstitutionFE + YearFE + \epsilon_{it}$$

This specification tests for nonlinear increasing or diminishing returns to concentration. Annual counterfactual elite output was calculated as:

$$CounterfactualElite_t = TotalElite_t \times ForeignFacultyShare_t$$

The elite surplus equals observed foreign elite output minus this proportional benchmark. Summation across years yields cumulative structural amplification.

The extended analyses clarify the multi-level structure of elite amplification. Logistic decomposition shows that approximately 54% of the raw elite gap is attributable to compositional sorting into high-impact domains and institutions. However, foreign-trained faculty retain a modest within-institution conditional advantage (≈3% higher odds of producing top-1% papers). The quantile gradient analysis reveals mild upper-tail steepening: the foreign effect approximately doubles between the top-10% and top-1% thresholds. The continuous FWCI model indicates a broader distributional shift, with foreign-trained faculty exhibiting approximately 7–8% higher normalized citation impact after controls.

Mixed-team analysis demonstrates strong complementarity: papers authored by both domestic- and foreign-trained faculty have approximately 30% higher odds of elite output than homogeneous teams. At the institutional level, elite output rises with foreign-trained faculty share but at a diminishing rate, suggesting steady rather than accelerating returns to concentration.

## S10. Sequential decomposition of the elite-output gap

To evaluate the structural sources of the foreign–domestic elite-output gap, we implemented a sequential Fairlie-style decomposition (28) for the binary top-1% citation outcome. The decomposition was estimated using grouped paper–faculty observations constructed from the linked AARC–OpenAlex panel spanning 2011–2020.

The outcome variable was a binary indicator equal to one if a publication belonged to the top-1% of field- and year-normalized citation impact. Because the outcome is binary, decomposition was implemented within a nonlinear generalized linear modeling framework rather than through a standard linear Oaxaca decomposition. We estimated a sequence of weighted logistic regression models of the form:

$$Pr(Y_{ijt} = 1) = f(Foreign_i, X_{ijt})$$

where:

- $Y_{ijt}$ denotes elite publication status for faculty–publication observation $i, j, t$,
- $Foreign_i$ indicates foreign degree training,
- $X_{ijt}$ denotes progressively incorporated structural covariates,
- and $f(\cdot)$ represents the logistic link function.

Publication year fixed effects were included in all specifications but were not themselves decomposed. Structural blocks were added sequentially in the following order:

1. disciplinary domain,
2. institutional research intensity,
3. institution type (public vs. private nonprofit),
4. academic rank/career stage.

For each specification, we computed the standardized foreign–domestic elite-output gap by predicting publication probabilities under counterfactual foreign and domestic assignments while holding the observed covariate distribution fixed. Specifically, predicted elite probabilities were generated twice for each observation: (1) setting all observations to foreign-degree status, (2) setting all observations to domestic-degree status, while preserving the observed structural covariates and publication-year composition. The weighted difference between the two sets of predicted probabilities defines the adjusted foreign–domestic elite-output gap for a given specification.

The sequential decomposition therefore measures how the remaining adjusted gap changes after progressively conditioning on structural characteristics of the research system. Negative changes indicate attenuation of the gap after introducing a structural block, whereas positive changes indicate modest widening due to overlapping or suppressing relationships among covariates. Because sequential decompositions in nonlinear models are order-dependent, the resulting quantities should be interpreted descriptively rather than causally. The analysis is intended to evaluate how much of the aggregate foreign–domestic elite-output gap is associated with structural concentration within specific disciplinary and institutional environments, rather than to estimate a causal treatment effect of foreign degree training itself.

## S11. Matched faculty-year comparison

Observations were retained for years 2011–2020 and required non-missing degree-origin information. Degree origin was coded as domestic if Degree country = “United States” and foreign otherwise. Faculty rank was assigned from the AARC rank files. Rank categories were reduced to three analytic levels: assistant, associate, and full professor.

For each faculty-year, we computed:

- Publications per year: distinct works associated with the faculty-year,
- Top 1% per year: number of works in the top-1% citation tier,
- Mixed team share: share of publications in mixed domestic–foreign teams,
- Degree centrality: number of distinct collaborators across all coauthored works,
- Corresponding author share: share of publications with corresponding-author status,

- Mean FWCI: average field-weighted citation impact across publications.

Team type at the work level was defined from the degree origin composition of coauthors: papers with only foreign-degree authors were coded “foreign_only”, papers with only domestic-degree authors were coded “domestic_only”, and all other papers were coded “mixed”. Degree centrality was computed on the coauthorship network implied by the faculty–publication file, counting distinct collaborators for each author across the full observation window.

To construct the matched comparison, we grouped faculty-year observations into strata defined by institution, disciplinary domain, rank, and year. For each outcome, we computed the mean value among foreign-trained faculty and the mean value among domestic-trained faculty within each stratum. Strata containing only one degree-origin group were excluded from the matched comparison because a within-stratum difference cannot be estimated there. The matched estimate for each outcome is the average of the within-stratum foreign–domestic differences across all valid strata. In the resulting analytic file, 38,223 strata contained both foreign- and domestic-trained faculty and contributed to the matched estimates.

### S12. Collaboration network analysis

Faculty collaboration networks were constructed using paper-level coauthorship data from the AARC-linked publication dataset. For each publication, pairs of U.S.-employed faculty authors were identified to generate an undirected faculty–faculty edge list. Self-loops were excluded, and duplicate edges were removed. The resulting network spans more than 12 million faculty–publication observations across 2011–2020.

Degree centrality was defined as the number of distinct AARC faculty collaborators per faculty member, computed by aggregating unique coauthor IDs across all publications. Cross-domain bridging was measured as the number of distinct umbrella domains represented among a faculty member’s collaborators. Cross-institution bridging was measured analogously using distinct collaborator institution identifiers.

For each faculty member, elite production was measured as the share of their publications classified as top-1% of field- and year-normalized citation impact. The correlation between degree centrality and elite probability was computed separately for foreign- and domestically trained faculty to assess whether network embeddedness translates differently into elite output across groups.

### S12. Topic Distinctiveness

Topic-level topic classifications were extracted from the OpenAlex primary_topic.id field for each publication (n = 4,516). Each paper was assigned a single topic identifier based on its primary topic.

For each year (2011–2020), we computed the frequency distribution of topics among all U.S.-employed faculty publications in the dataset. Annual topic probabilities were defined as:

$$P(topic \mid year) = \frac{\#publications\ in\ topic\ at\ year\ t}{\#total\ publications\ at\ year\ t}$$

This annual baseline controls for macro-level intellectual drift and evolving research trends.

For each author-year observation, we computed a topic distinctiveness index defined as:

$$Distinctiveness_{it} = \sum_{s} P(s|i,t)(-logP(s \mid U.S.,t))$$

where:

- $s$ indexes topics,
- $P(s|i,t)$ is the share of author *i*'s publications in topic *s* in year *t*,
- $P(s \mid U.S.,t)$ is the U.S.-wide publication share in topic *s* in year *t*.

In the primary-topic specification, this reduces to:

$$Distinctiveness_{it} = -\log(P(topic_{i,t} \mid U.S.,t))$$

Higher values indicate publication in less common topics relative to the annual U.S. baseline.

The topic distinctiveness analysis does not support the hypothesis that foreign-trained faculty are more likely to publish in rarer intellectual topics. Across all academic ranks, foreign-trained faculty exhibit slightly lower average distinctiveness scores than domestically trained faculty. The consistency of this pattern across assistant, associate, and full professor ranks indicates that topical differentiation is neither pronounced at entry nor attenuated at higher levels of institutional seniority.